\documentclass[aps,prb,twocolumn,groupedaddress,showpacs,10pt]{revtex4-1}
\usepackage{amsmath}
\usepackage{amssymb}
\usepackage{bm}
\usepackage[utf8]{inputenc}
\usepackage{mathptmx}
\usepackage{graphicx}
\usepackage{multirow}
\usepackage{lmodern}
\usepackage{color}
\usepackage{units}
\usepackage{dcolumn}
\usepackage{comment}
\usepackage[bookmarks=true,colorlinks=true,linkcolor=blue,citecolor=blue,urlcolor=blue]{hyperref}
\usepackage[figure]{hypcap}

\begin{document}

% Use the \preprint command to place your local institutional report
% number in the upper righthand corner of the title page in preprint mode.
% Multiple \preprint commands are allowed.
% Use the 'preprintnumbers' class option to override journal defaults
% to display numbers if necessary
%\preprint{}

%Title of paper
\title{Weak ferromagnetism in hexagonal Mn$_3$Z (Z=Sn, Ge, Ga) alloys}

% repeat the \author .. \affiliation  etc. as needed
% \email, \thanks, \homepage, \altaffiliation all apply to the current
% author. Explanatory text should go in the []'s, actual e-mail
% address or url should go in the {}'s for \email and \homepage.
% Please use the appropriate macro foreach each type of information

% \affiliation command applies to all authors since the last
% \affiliation command. The \affiliation command should follow the
% other information
% \affiliation can be followed by \email, \homepage, \thanks as well.
\author{B. Ny\'ari}
\author{A. De\'ak}  % ORCID: https://orcid.org/0000-0002-3210-2947
\author{L. Szunyogh}
\affiliation{Department of Theoretical Physics, Budapest University of Technology and Economics, Budafoki \'ut 8., HU-1111 Budapest, Hungary}
\affiliation{MTA-BME Condensed Matter Research Group, Budapest University of Technology and Economics, Budafoki \'{u}t 8., HU-1111 Budapest, Hungary} 

\date{\today}

\begin{abstract}
We present combined spin model and first principles electronic structure calculations to study the weak ferromagnetism in bulk Mn$_3$Z (Z=Sn, Ge, Ga) compounds. The spin model parameters were determined from a spin-cluster expansion technique based on the relativistic disordered local moment formalism implemented in the screened Korringa--Kohn--Rostoker method. We describe the magnetic ground state of the system within a three-sublattice model and investigate the formation of the weak ferromagnetic states in terms of the relevant model parameters. First, we give a group-theoretical argument how the point-group symmetry of the lattice leads to the formation of weak ferromagnetic states. Then we study the ground states of the classical spin model and derive analytical expressions for the weak ferromagnetic distortions by recovering the main results of the group-theoretical analysis. As a third approach we obtain the weak ferromagnetic ground states from self-consistent density functional calculations and compare our results with previous first principles calculations and with available experimental data. In particular, we demonstrate that the orbital moments follow a decomposition predicted by group theory. For a deeper understanding of the formation of weak ferromagnetism we selectively trace the effect of the spin-orbit coupling at the Mn and Z sites. In addition, for the case of Mn$_3$Ga, we gain information on the role of the induced moment of Ga from constrained local density functional calculations.
\end{abstract}

% insert suggested PACS numbers in braces on next line
%\pacs{75.10.Hk, 71.15.Mb, 71.15.Rf}
% 75.70.Ak 	Magnetic properties of monolayers and thin films
% 71.15.Mb 	Density functional theory, local density approximation, gradient and other corrections 
% 71.15.Rf 	Relativistic effects
% ? 75.10.Hk  Classical spin models?

%\maketitle must follow title, authors, abstract, \pacs, and \keywords
\maketitle

% body of paper here - Use proper section commands

% If in two-column mode, this environment will change to single-column
% format so that long equations can be displayed. Use
% sparingly.
%\begin{widetext}
% put long equation here
%\end{widetext}

% Surround figure environment with turnpage environment for landscape
% figure
% \begin{turnpage}
% \begin{figure}
% \includegraphics{}%
% \caption{\label{}}
% \end{figure}
% \end{turnpage}

% The ruledtabular enviroment adds doubled rules to table and sets a
% reasonable default table settings.
% Use the table* environment to get a full-width table in two-column
% Add \usepackage{longtable} and the longtable (or longtable*}
% environment for nicely formatted long tables. Or use the the [H]
% placement option to break a long table (with less control than 
% in longtable).
% \begin{table}%[H] add [H] placement to break table across pages
% \caption{\label{}}
% \begin{ruledtabular}
% \begin{tabular}{}
% Lines of table here ending with \\
% \end{tabular}
% \end{ruledtabular}
% \end{table}

%% Create the reference section using BibTeX:
%\bibliography{basename of .bib file}

%end of comments

\section{Introduction}\label{intro}

The Mn$_3$Z (Z=Sn, Ge, Ga) compounds, known to be weak ferromagnetic at low temperatures\cite{tomiyoshi-1982,tomiyoshi-1983,kren-1970}, are in the focus of current research interest due to the exotic topological properties of their band structure. They show large anomalous Hall conductivity due to the nonvanishing Berry curvature induced by the non-collinear triangular ground state spin structure\cite{Kubler-2014,nakatsuji-2015,nayak-2016} and are %also possible candidates for 
topological Weyl semimetals because of emerging Weyl nodes in the band structure near the Fermi level\cite{kubler-2017,Yang-2017}. 
The Mn$_3$Z compounds are also possible candidates to replace the expensive IrMn based antiferromagnets in magnetic sensors based on the GMR effect\cite{hirohata-2017}.

\begin{figure}[htb]
\includegraphics[width=0.9\linewidth]{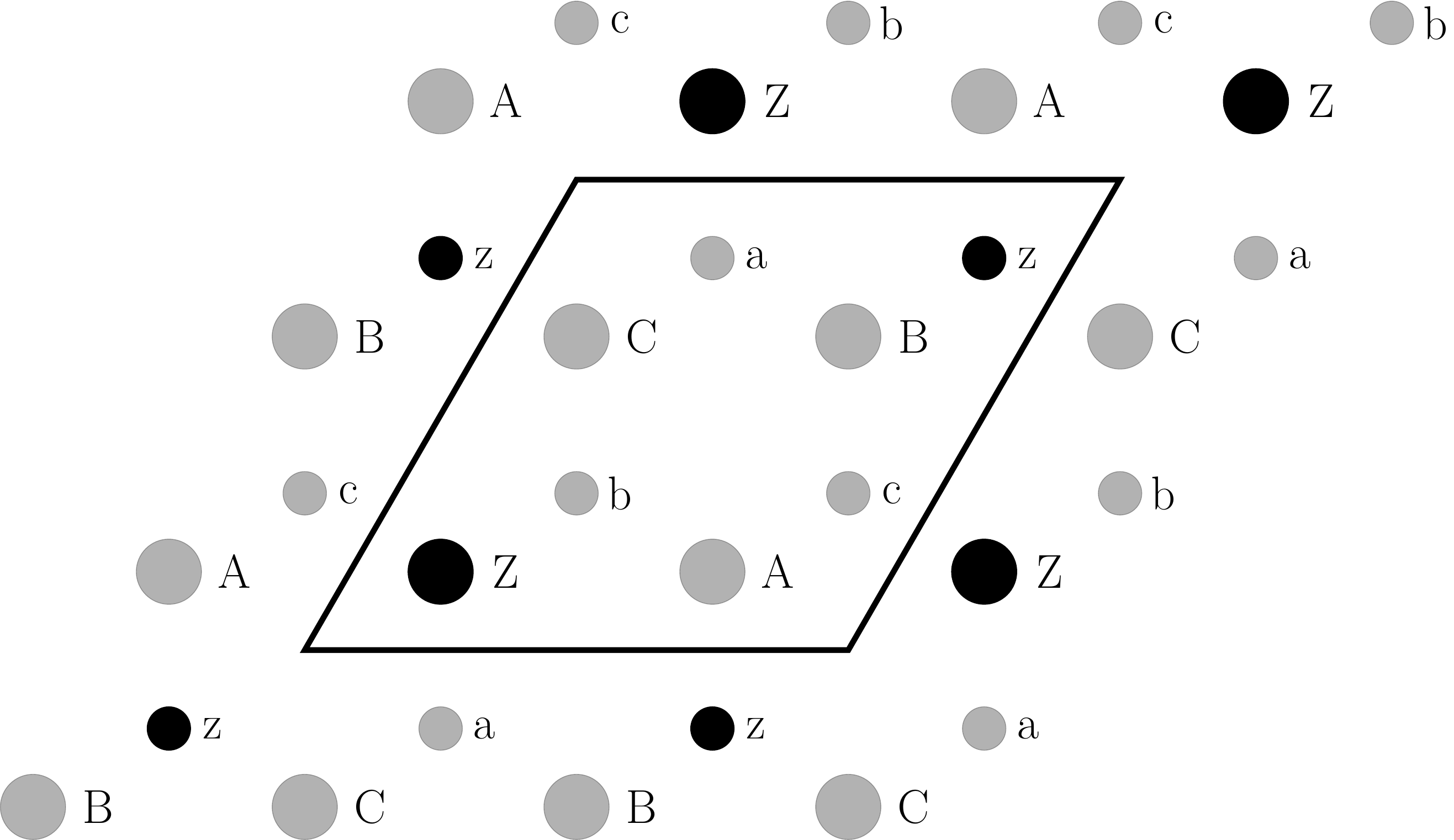}
\caption{The atomic positions in the hexagonal phase of the Mn$_3$Z  compounds: large and small circles, labelled by capital and small letters, denote sites in the atomic layers at $z=c/4$ and at $z=3c/4$, respectively, while grey and black circles stand in order for Mn and Z atoms. The rhombus encloses a possible unit cell of the system.}
\label{fig:structure}
\end{figure}

The Mn$_3$Z  compounds have three different structural phases: a hexagonal phase with DO$_{19}$ structure, a tetragonal phase with DO$_{22}$ structure and a cubic phase with a standard Heusler structure. These phases and the transition between them have been the subject of recent research\cite{zhang-2013,khmelevskyi-2016}. In this paper, we are going to investigate the magnetic properties in the hexagonal phase of these compounds.
The atomic positions in the DO$_{19}$ hexagonal phase are sketched in Fig.~\ref{fig:structure}. The Mn sites in each layer form a kagome lattice, i.e.\ a two-dimensional network of corner-sharing equilateral triangles. The atomic layers are shifted alternately with $\frac{2}{3}\left(\vec{a}+\vec{b}\right)$ and $\frac{1}{3}\left(\vec{a}+\vec{b}\right)$ where $\vec{a}$ and $\vec{b}$ are the primitive vectors of the kagome lattice. The unit cell marked by the black rhombus in Fig.~\ref{fig:structure} contains two layers built up from six manganese atoms and two non-magnetic Z atoms.

\begin{figure}[htb]
\includegraphics[width=\linewidth]{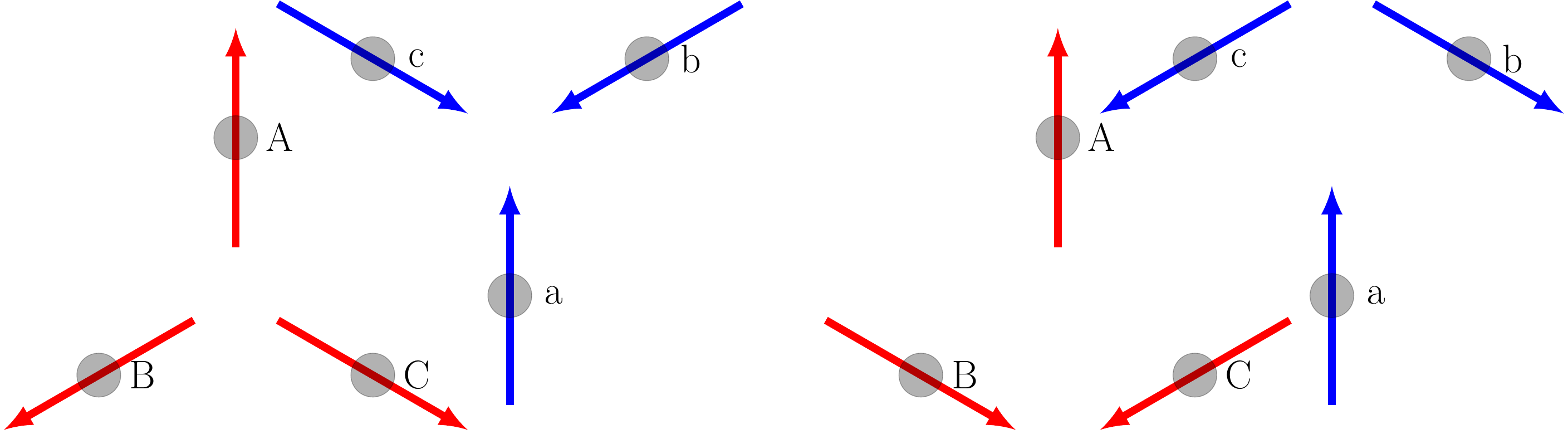}

\medskip
\ $\Gamma_3$   \hskip 4.5cm       $\Gamma_5$ 
\caption{Low-energy chiral magnetic structures of the Mn$_3$Z compounds. % the state on the left is denoted by $\Gamma_3$, while the one on the right by $\Gamma_5$.
}
\label{fig:spin_structure}
\end{figure}

The magnetic structure of these materials was measured with polarized neutron diffraction experiments showing that the low-energy magnetic states of these compounds in the hexagonal phase are chiral antiferromagnetic (AFM) states\cite{tomiyoshi-1982,tomiyoshi-1983,kren-1970} as illustrated in Fig.~\ref{fig:spin_structure}, with a small distortion producing a tiny net magnetic moment. A simple spin-model analysis in Ref.\ \onlinecite{tomiyoshi-1982} proved the magnetocrystalline anisotropy to be the microscopic mechanism responsible for the weak ferromagnetic (WF) distortion, whereas the Dzyaloshinsky--Moriya (DM) interaction \cite{dzyaloshinskii-1958,moriya-1960} was shown to lift the chiral degeneracy of the $\Gamma_3$ and $\Gamma_5$ states.  

The electronic and magnetic structure of these compounds was also investigated theoretically in terms of self-consistent field density functional calculations in the local density approximation (LDA) \cite{sandratskii-1996,Kubler-2014,zhang-2013,khmelevskyi-2016,mendive-tapia-2019}. In the hexagonal phase, the chiral $\Gamma_5$ state with a weak ferromagnetic (WF) distortion and a small net magnetic moment was found as ground state \cite{sandratskii-1996,Kubler-2014,zhang-2013}.  In particular, Ref.\ \onlinecite{sandratskii-1996} discussed symmetry considerations and the role of orbital polarization on the formation of weak ferromagnetism in Mn$_3$Sn. 
It should be noted that in Ref.\ \onlinecite{zhang-2013} the tetragonal phase of Mn$_3$Sn was found lower in energy than the hexagonal phase and the stability of the hexagonal phase according to experiments was attributed to structural disorder or off-stoichiometric compositions. The influence of these effects on the magnetic ordering in the Mn$_3$Ga alloy has been studied in Ref.\ \onlinecite{khmelevskyi-2016}.
                               
In this work we present a detailed theoretical investigation of the magnetic ground state of the Mn$_3$Z  alloys in the hexagonal phase. We employ the relativistic Screened Korringa--Kohn--Rostoker (SKKR) method\cite{szunyogh-1994,zeller-1995} to calculate the electronic structure and the magnetic properties. In particular, we set up a classical spin model with parameters obtained from the combination of the spin-cluster expansion (SCE) and the relativistic disordered local moment (RDLM) method\cite{szunyogh-2011}. Using the point-group symmetry of the lattice we determine the general parametric form of the exchange interaction matrices of a three-sublattice model and we provide the group theoretical reason behind the formation of the weak ferromagnetic state. By solving the spin model we quantify the weak ferromagnetic distortion in terms of the model parameters. Our results are clearly consistent with the original spin-model description of weak ferromagnetism in Mn$_3$Sn \cite{tomiyoshi-1982}, however, we exceed this approach by quantitative estimates on the weak ferromagnetic distortion being in fairly good agreement with the experiments.  We also obtain the magnetic ground states of the Mn$_3$Z compounds from unconstrained self-consistent LDA calculations and investigate 
the effect of spin-orbit coupling on the Mn and Z sites selectively. Finally, we give a hint to the effect of the induced moment at the Z site by performing constrained LDA calculations for the case of the Mn$_3$Ga alloy.  

\section{Methods}\label{methods}
\subsection{Spin model}\label{sm}
In order to study the magnetic properties of the Mn$_3$Z alloys in the hexagonal phase we use a classical Heisenberg model for the Mn spins represented with a set of unit vectors $\{\vec{e}\}$ and neglect the effect of the induced spin moment on the Z sites. The spin model in second order of the spin variables is given by 
\begin{equation}
H(\{\vec{e}\}) = \sum_{i} \vec{e}_i \mathbf K_{i} \vec{e}_i -\dfrac{1}{2}\sum_{i\neq j}\vec{e}_i \mathbf J_{ij} \vec{e}_j\;,
\label{Eq:Heisenberg}
\end{equation}
where the $i$ and $j$ indices are confined to the Mn sites, $\mathbf K_{i}$ are the second order on-site anisotropy matrices and $\mathbf J_{ij}$ are the tensorial exchange couplings. The exchange matrix can be decomposed as
\begin{equation}
\mathbf J_{ij} = J_{ij}\mathbf I + \frac{1}{2} \left( \mathbf J_{ij} - \mathbf J_{ij} ^T  \right) + \frac{1}{2} \left( \mathbf J_{ij} + \mathbf J_{ij} ^T - 2 J_{ij}\mathbf I \right)\; ,
\end{equation}
where  $\mathbf I$ is the unit matrix and $T$ denotes the transpose of a matrix. In the above decomposition $J_{ij} = \frac{1}{3}\mathrm{Tr} \mathbf J_{ij} $ defines the isotropic Heisenberg coupling between two spins, the antisymmetric part of the exchange tensor can be related to the Dzyaloshinsky--Moriya interaction,
\begin{equation}
\vec{e}_i\frac{1}{2}\left( \mathbf J_{ij} - \mathbf J_{ij} ^T  \right)\vec{e}_j = \vec{D}_{ij} \left( \vec{e}_i \times \vec{e}_j \right),
\end{equation}
and the traceless symmetric part of $\mathbf J_{ij}$ corresponds to the two-site anisotropy. 

%\subsection{Sublattice model}\label{sublattice}

From previous experimental\cite{tomiyoshi-1982,tomiyoshi-1983,kren-1970} and theoretical\cite{sticht-1989,sandratskii-1996} works it turns out that the ground state magnetic structure of the Mn$_3$Z compounds can be well described in terms of an effective spin model related to three Mn sublattices. This means that the A--a, B--b and C--c sublattice pairs (cf.\ Fig.\ \ref{fig:structure}) are strongly coupled ferromagnetically, ensuring that the corresponding Mn moments are parallel to each other. Consequently we only have to consider three independent sublattices to explore the low-energy magnetic configurations. As will be shown in Sec.\ \ref{sec:spinmodel} our calculated exchange interactions clearly support this observation, which leads to the following simplified Hamiltonian:
\begin{equation}
H = -\dfrac{1}{2}\sum_{\alpha,\beta=1}^{3}\vec{e}_\alpha \mathbf J_{\alpha\beta} \vec{e}_\beta\;,
\label{eq:H_sublatt}
\end{equation}
where the $\mathbf J_{\alpha\beta}$ matrices are the effective sublattice interactions. The $\mathbf J_{\alpha\beta}$ matrices can be related to the exchange matrices in Eq.\ \eqref{Eq:Heisenberg} as
\begin{equation}
\mathbf J_{\alpha\beta} = \sum_{n} \mathbf J_{0\alpha,n\beta} -2 \delta_{\alpha\beta} \mathbf K_\alpha \, ,
\label{eq:sum_sublattice}
\end{equation}
where the index $0$ stands for a fixed site in sublattice $\alpha$,  while $n$ goes through the sites in sublattice $\beta$. Note that due to translation invariance the on-site anisotropy matrices at all sites in a given sublattice are identical, which explains the notation $\mathbf K_\alpha$ in Eq.\ \eqref{eq:sum_sublattice}. 
%\begin{comment}
By collecting the spin variables of the three sublattices  into a nine-dimensional composite variable, $ \vec{e}=( \vec{e}_1,  \vec{e}_2,  \vec{e}_3)$ and the sublattice interactions into a composite matrix,
\begin{equation}
\mathbf J = \left(
\begin{array}{ccc}
\mathbf J_{11} & \mathbf J_{12} & \mathbf J_{13} \\
\mathbf J_{21} & \mathbf J_{22} & \mathbf J_{23} \\
\mathbf J_{31} & \mathbf J_{32} & \mathbf J_{33} 
\end{array}
\right) \, ,
\label{eq:Jmat}
\end{equation}
Eq.~\eqref{eq:H_sublatt} can be rewritten into the simple form,
\begin{equation}
H = -\dfrac{1}{2} \vec{e} \,  \mathbf J \, \vec{e}\; .
\label{eq:Hcomp}
\end{equation}
%\end{comment}

The structure of the matrices \eqref{eq:sum_sublattice} can be obtained by using the $D_{3h}$ point-group symmetry of the lattice. This is provided by the invariance of the energy of the spin system \eqref{eq:H_sublatt} against any point-group element $g \in D_{3h}$. Denoting the $9\times9$ matrix representation of $g$ by $\mathbf R_g$, this implies the relationships,
\begin{equation}
\mathbf J =  \mathbf R_g^T \mathbf J \mathbf R_g\;.
\label{eq:symmetry}
\end{equation}
\begin{comment}
\begin{equation}
\mathbf J_{\alpha^\prime  \beta^\prime} =  \mathbf R_g^T \mathbf J_{\alpha \beta} \mathbf R_g\;.
\label{eq:symmerty}
\end{equation}
\end{comment}
Note that the permutation of the sublattices under the operation $g$ is also included in the representation $\mathbf R_g$.
%$\alpha^\prime$ stands for the index of the sublattice, the sublattice $\alpha$ is transformed to under the action of the inverse of $g$.
Performing the corresponding analysis we obtain two different types of sublattice interaction matrices: the three sublattice-diagonal matrices are connected via the $C_3$ rotation, while the six sublattice off-diagonal matrices are related to each other either by $C_3$ rotation or by transposition. One representative element for each set is given by
\begin{equation}
\mathbf J_{AA} = 
\left(
\begin{array}{ccc}
\frac{-K^{x} + 3K^{y}}{2} & 0 & 0 \\
0 & \frac{3K^{x} -  K^{y}}{2} & 0 \\ 
0 & 0 & -K^{x}-K^{y}
\end{array}
\right)
\label{eq:JAA}
\end{equation} 
and
\begin{equation}
\mathbf J_{BC} = 
\left(
\begin{array}{ccc}
J+J^{x} & D & 0 \\- D & J+J^{y} & 0 \\0 & 0 & J-J^{x}-J^{y}
\end{array}
\right) \, ,
%+J\mathbf I\;,
\label{eq:JBC}
\end{equation} 
respectively.
The sublattice model has therefore six independent parameters: two sublattice-diagonal anisotropy constants, $K^{x}$ and $K^{y}$, two sublattice-off-diagonal anisotropy constants, $J^{x}$ and $J^{y}$, one Dzyaloshinsky--Moriya parameter, $D$, describing an effective DM vector parallel to the $z$ axis, and an isotropic coupling between different sublattices, $J$. Note that in principle there is an isotropic coupling parameter for the sublattice-diagonal matrices, but it only adds a constant to the energy, thus it has no effect on the magnetic ordering in the system.  

\subsection{\emph{Ab initio} calculations} \label{abinitio}
We performed self-consistent electronic structure calculations for the Mn$_3$Z  compounds in terms of the relativistic Screened Korringa--Kohn--Rostoker method\cite{szunyogh-1994,zeller-1995}.  
The lattice constants of the different compounds were set to the experimental values shown in Table~\ref{table:lattice}.

\begin{table}[htb]
\caption{The experimental lattice constants for the Mn$_3$Z  compounds}
\def\arraystretch{1.2}
\begin{ruledtabular}
    \begin{tabular}{c|ccc}
                 & Mn$_3$Sn\cite{zimmer-1971} & Mn$_3$Ge\cite{kadar-1971} & Mn$_3$Ga\cite{kren-1970}\\ \colrule
      a$_\text{2d}$[\AA] & 5.665   & 5.36   & 5.36  \\
      $c/a$      & 0.79982 & 0.80598& 0.807
    \end{tabular}
    \label{table:lattice}
\end{ruledtabular}
\end{table}

We used the local spin-density approximation parametrized according to Vosko \emph{et al.},\cite{vosko-1980} and we employed the atomic sphere approximation with an angular momentum cutoff of $\ell_\text{max}=2$. We used 16 energy points on a semicircular path on the upper complex half-plane for the energy integrations, and 144 points in the 2D Brillouin zone (2DBZ) for $k$-integrations. %Note that we couldn't utilize the lattice symmetry due to the lower symmetry of the weak ferromagnetic states. 

In order to obtain the parameters of the tensorial Heisenberg model \eqref{Eq:Heisenberg}, we employed the spin-cluster expansion developed originally by Drautz and F\"ahnle\cite{drautz-2004,drautz-2005} combined with the relativistic disordered local moment method \cite{gyorffy-1985, staunton-2004,staunton-2006}. The RDLM method provides a first-principle description of a paramagnetic system according to the adiabatic decoupling of the electronic and spin degrees of freedom, while the SCE enables a systematic parametrization of the adiabatic energy surface. For the details of the SCE-RDLM method see Ref.~\onlinecite{szunyogh-2011}. In the SCE calculations we used 16 energy points on a semicircular path on the upper complex half-plane with approximately 20000 $k$ points in the 2DBZ near the Fermi energy.  

\section{Results}\label{sec:results}

\subsection{Spin model parameters}\label{sec:spinmodel}

\begin{figure}[htb]
\includegraphics[width=0.9\linewidth]{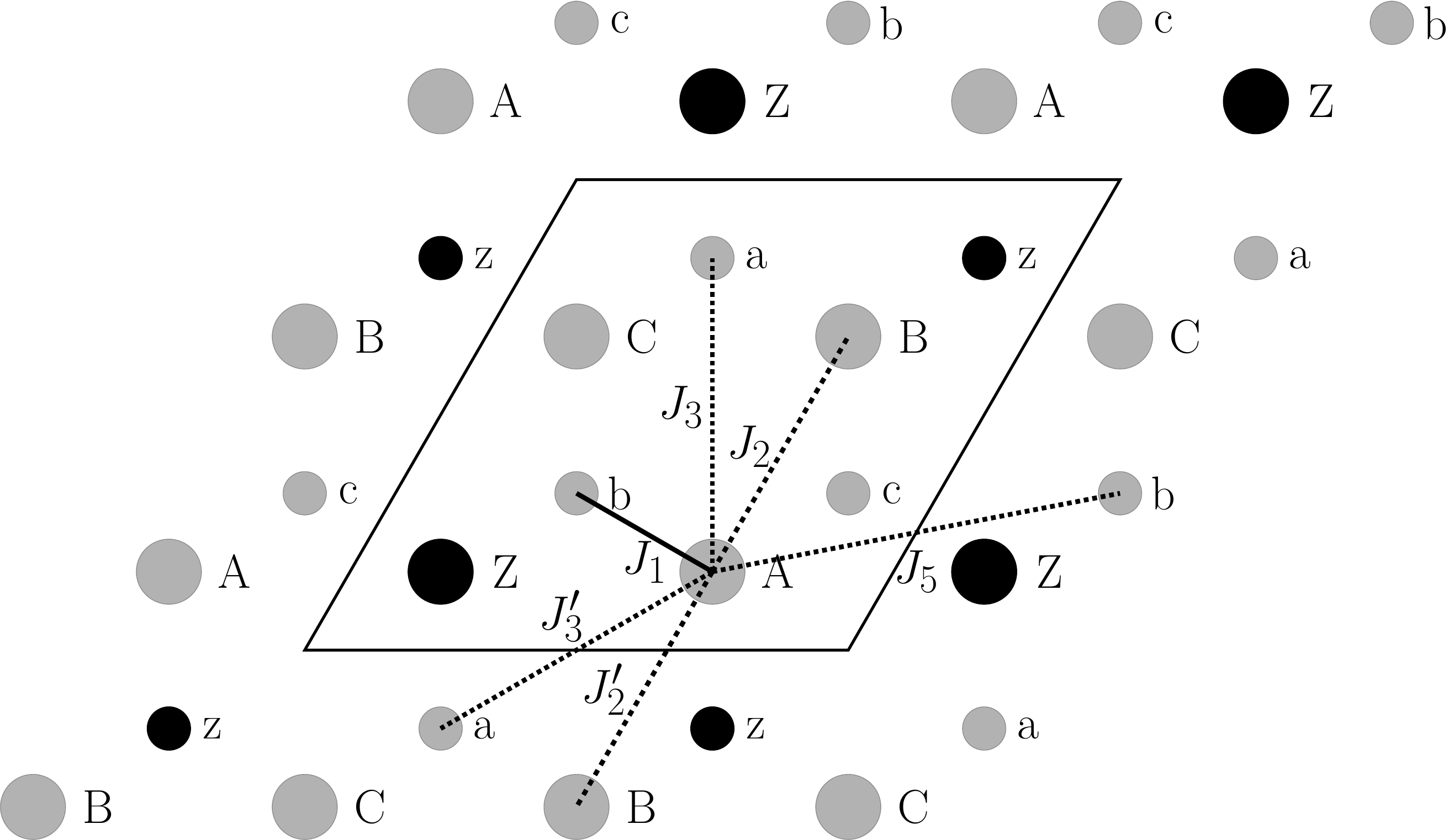}
\caption{Schematic view of the first five nearest neighbor interactions. Different interactions for the same distance are denoted with and without primes. The $J_4$ interaction connects two sites in the same sublattice from neighboring unit cells shifted along the $z$ axis, thus, we could not illustrate it in the figure.}
\label{fig:isotrop_int}
\end{figure}

We applied the SCE-RDLM method to obtain the \emph{ab initio} spin model parameters in the paramagnetic phase of each compound. First, we discuss the isotropic couplings. In Table \ref{table:isotropic} we show the first five nearest neighbor interactions as visualized in Fig.~\ref{fig:isotrop_int}. For all the three compounds we find a similar structure of the isotropic interactions. The nearest neighbor interactions $J_1$ couple sites that belong to different sublattices and different atomic layers. They are antiferromagnetic (AFM) and the largest in magnitude among the isotropic couplings.  By contrast, the third nearest neighbor out-of-plane interactions,  $J_3$ and $J_3^\prime$, that connect sites in the same sublattice are strongly ferromagnetic.  These interactions thus force to align the moments in the same sublattice irrespective of the atomic-layer positions.  On top of this, the AFM first nearest neighbor couplings cause frustration on the kagome lattice and stabilize the triangular states as shown in Fig.\ \ref{fig:spin_structure}. The (in-plane) second nearest neighbor antiferromagnetic interactions, $J_2$, also contribute to the stabilization of the triangular spin structures, though the other type of these interactions, $J_2^\prime$,
is ferromagnetic, destabilizing the triangular state to some extent. By using the {\em ab initio} tensorial spin model we solved the Landau--Lifshitz--Gilbert equations at zero  temperature (with damping term only) and for all the three compounds we indeed obtained a ground state close to the $\Gamma_5$ state. As we will demonstrate later, it is the Dzyaloshinsky--Moriya interaction which selects between the $\Gamma_3$ and $\Gamma_5$ spin states.

\begin{table}[htb]
    \caption{Calculated isotropic couplings in the Mn$_3$Z  compounds. The interactions given in units of meV are indexed according to increasing distances of the pairs, while interactions with and without prime stand for inequivalent pairs with the same distance (see Fig.~\ref{fig:isotrop_int}). For better understanding, in the second row the in-plane and out-of-plane couplings are denoted by ip and oop, respectively.}
    \label{table:isotropic}
    \begin{ruledtabular}
        \begin{tabular}{c|rrrrrrr}
                 & \multicolumn{1}{c}{$J_1$} & \multicolumn{1}{c}{$J_2$} &\multicolumn{1}{c}{$J_2^{\prime}$}& \multicolumn{1}{c}{$J_3$} & \multicolumn{1}{c}{$J_3^{\prime}$}& \multicolumn{1}{c}{$J_4$} & \multicolumn{1}{c}{$J_5$} \\ 
                 & \multicolumn{1}{c}{oop}   & \multicolumn{1}{c}{ip}    &\multicolumn{1}{c}{ip}         & \multicolumn{1}{c}{oop}  & \multicolumn{1}{c}{oop}           & \multicolumn{1}{c}{oop}  & \multicolumn{1}{c}{oop}   \\\colrule
        Mn$_3$Sn & -15.3 & -3.2  & 4.5          & 13.8  & 11.3          &  -2.87& -4.08 \\
        Mn$_3$Ge & -22.6 & -7.4  & 10.1         & 5.9   & 7.1           &  -2.54& -4.39 \\
        Mn$_3$Ga & -23.7 & -15.9 & 0.9          & 10.7  & 5.5           &  -3.82& -4.42
        \end{tabular}
    \end{ruledtabular}
\end{table}

\begin{figure}[htb]                               
\includegraphics[width=\linewidth]{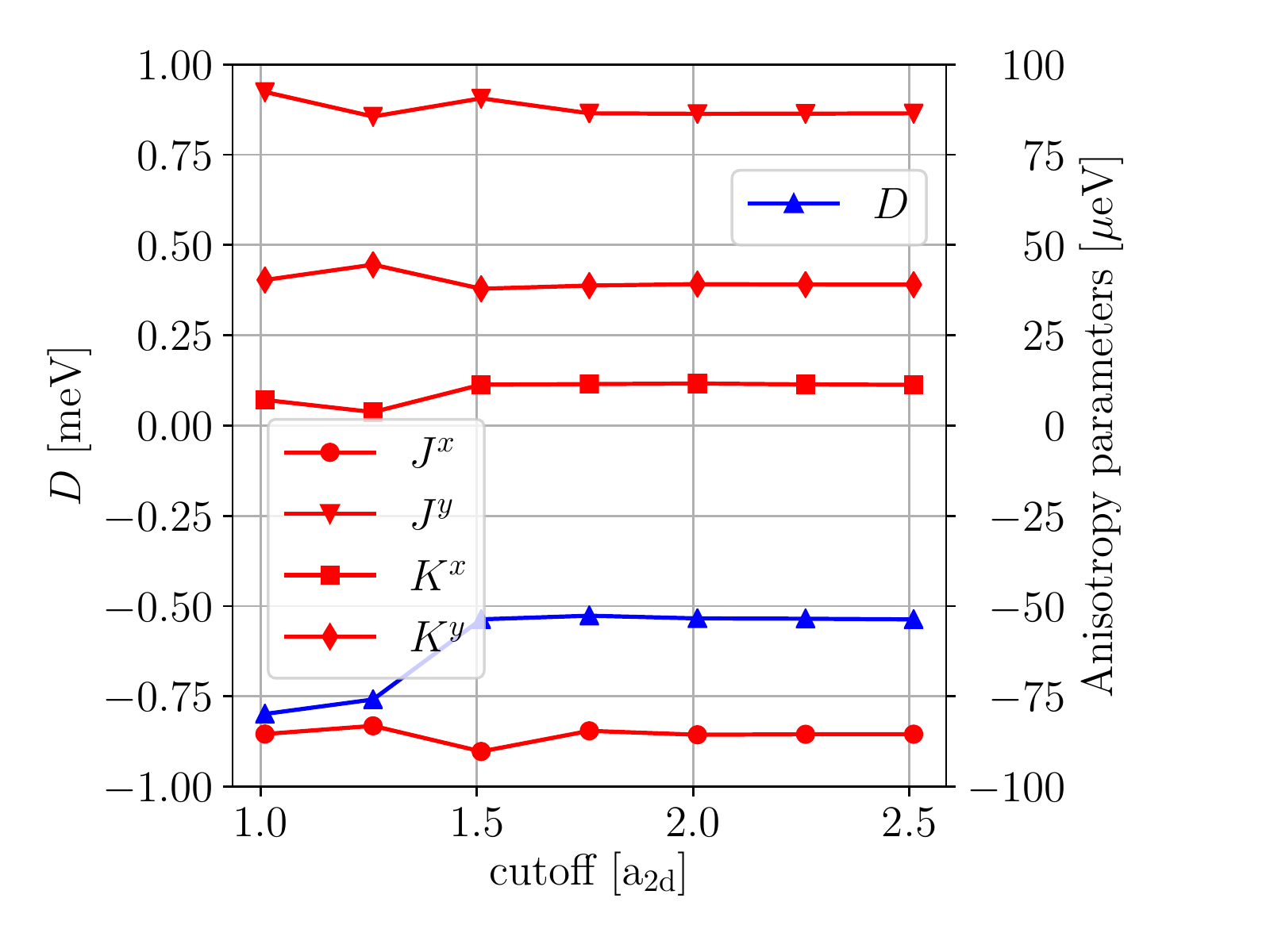}
\vskip -10pt
\caption{The cutoff dependence of the relativistic interactions in the sublattice  matrices \eqref{eq:JAA} and \eqref{eq:JBC} for Mn$_3$Sn. Note that the DM parameter $D$ is about one order of magnitude larger than the anisotropy parameters. The isotropic coupling $J$ (not presented here) shows a similar cutoff dependence.}
\label{fig:cutoff}
\end{figure}

From the $\mathbf{J}_{ij}$ and $\mathbf{K}_i$ matrices we can calculate the $\mathbf{J}_{\alpha\beta}$ matrices as defined in Eq.~\eqref{eq:sum_sublattice}. We checked that the obtained matrices satisfy to high accuracy the analytic forms  \eqref{eq:JAA} and \eqref{eq:JBC} we deduced from symmetry principles, so the six parameters of the sublattice model can be read off.  The values of these parameters depend on the cutoff distance of pairs in the sum in Eq.~\eqref{eq:sum_sublattice}. The dependence of the DM and the anisotropy parameters for Mn$_3$Sn is shown in Fig.~\ref{fig:cutoff}. Clearly, all the parameters converge well beyond a distance of about $2 \,a_{\text{2d}}$. 
The other two compounds show a similar behavior. Based on these results, in all cases we used a cutoff of $2.51\, a_{\text{2d}}$ for the calculation of the sublattice model parameters. 

\begin{table}[htb]
    \caption{Calculated sublattice model parameters, see Eqs.~\eqref{eq:JAA} and \eqref{eq:JBC},  for the Mn$_3$Z  compounds based on the SCE-RDLM method.}
    \label{table:sublattice_params}
    \begin{ruledtabular}
        \begin{tabular}{c|cccccc}
		& $J$[meV] & $D$[meV] & $K^x$[$\mu$eV]&  $K^y$[$\mu$eV] &  $J^x$[$\mu$eV] & $J^y$[$\mu$eV]  \\\colrule
	Mn$_3$Sn & -46.7 & -0.547 &  11 &  39 & -85 & 86 \\ 
	Mn$_3$Ge & -51.6 & -0.246 & -64 &123 &-108 & 92 \\ 
	Mn$_3$Ga & -77.0 & -0.447 & -81 &129 & -65 &135 \\ 
	\end{tabular}
    \end{ruledtabular}
\end{table}

The calculated parameters of the sublattice model, see Eqs.~\eqref{eq:JAA} and \eqref{eq:JBC}, are summarized in Table~\ref{table:sublattice_params}. In each case a large antiferromagnetic isotropic coupling was obtained, which again explains  the formation of the low-energy frustrated triangular configurations. The DM parameter being two orders of magnitude less than $J$ has negative sign, thus, it prefers the $\Gamma_5$ state against the $\Gamma_3$  state for all of the compounds. The anisotropy constants are typically one order of magnitude less than the DM parameters. Remarkably, the anisotropy parameters indexed by $x$ are positive and those indexed by $y$ are negative. The only exception is observed for $K^x$ in case of Mn$_3$Sn, which is negative in sign.

\subsection{Group-theoretical argument}\label{grouptheory}

\begin{figure}[htb]
%\begin{ruledtabular}
    \begin{tabular}{|c|c|}
    \colrule
      \quad \rotatebox[origin=c]{90}{$A^{\prime}_1 = \Gamma_{3,y}$} \quad
          & \begin{minipage}[c]{.85\linewidth}
	       \vspace*{0.2cm}
	       \includegraphics[width= .3\linewidth]{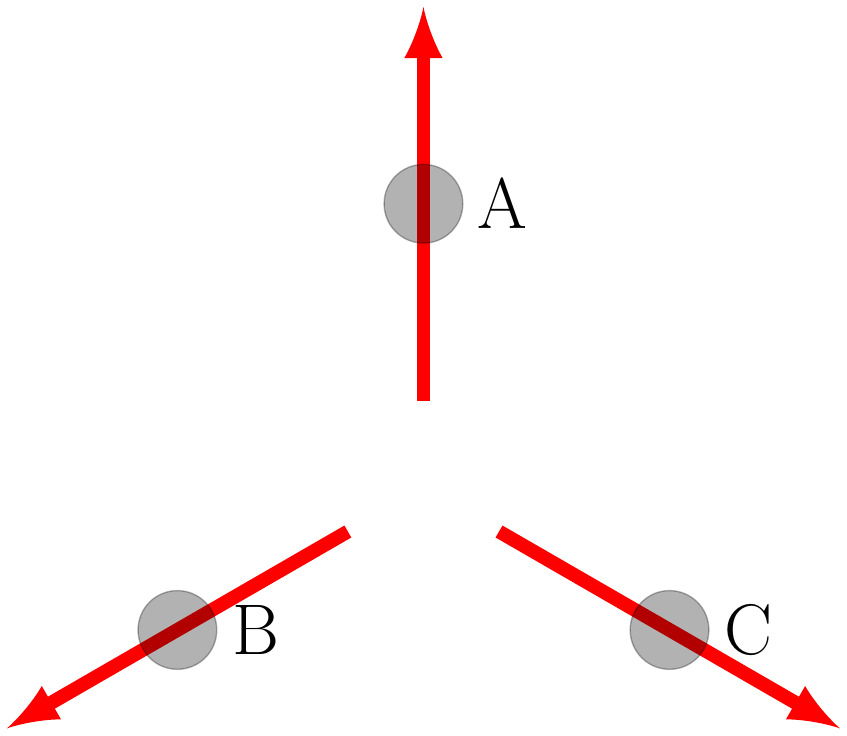}
	       \vspace*{0.3cm}
	    \end{minipage} \\
          \colrule 
      \quad \rotatebox[origin=c]{90}{$A^{\prime}_2 = \Gamma_{3,x}$} \quad
          & \begin{minipage}[c]{.85\linewidth}
	       \vspace*{0.3cm}
	       \includegraphics[width= .25\linewidth]{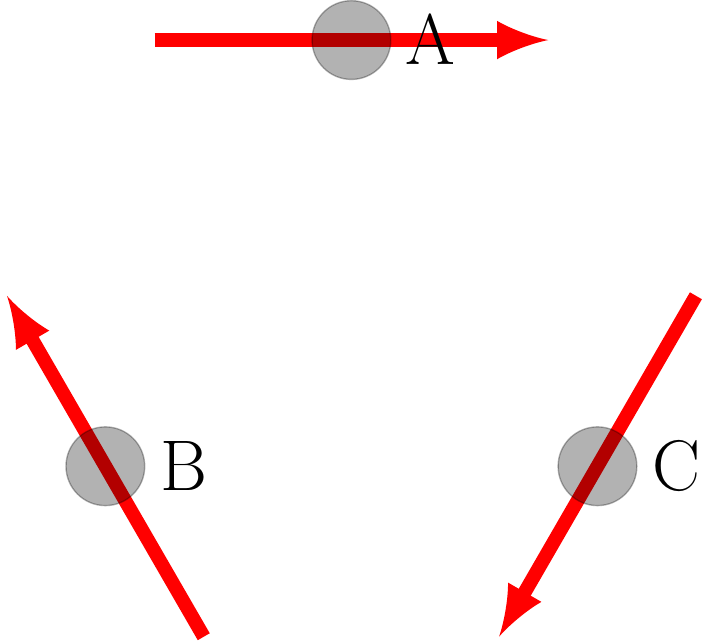}
	       \vspace*{0.3cm}
	    \end{minipage}  \\
	  \colrule
      \quad \rotatebox[origin=c]{90}{$2E^{\prime} =\{ \Gamma_{5,x},\text{FM}_x,\Gamma_{5,y},\text{FM}_y\}$} \quad
          & \begin{minipage}[c]{.85\linewidth}
	       \vspace*{0.3cm}
	       \includegraphics[width=  0.75\linewidth]{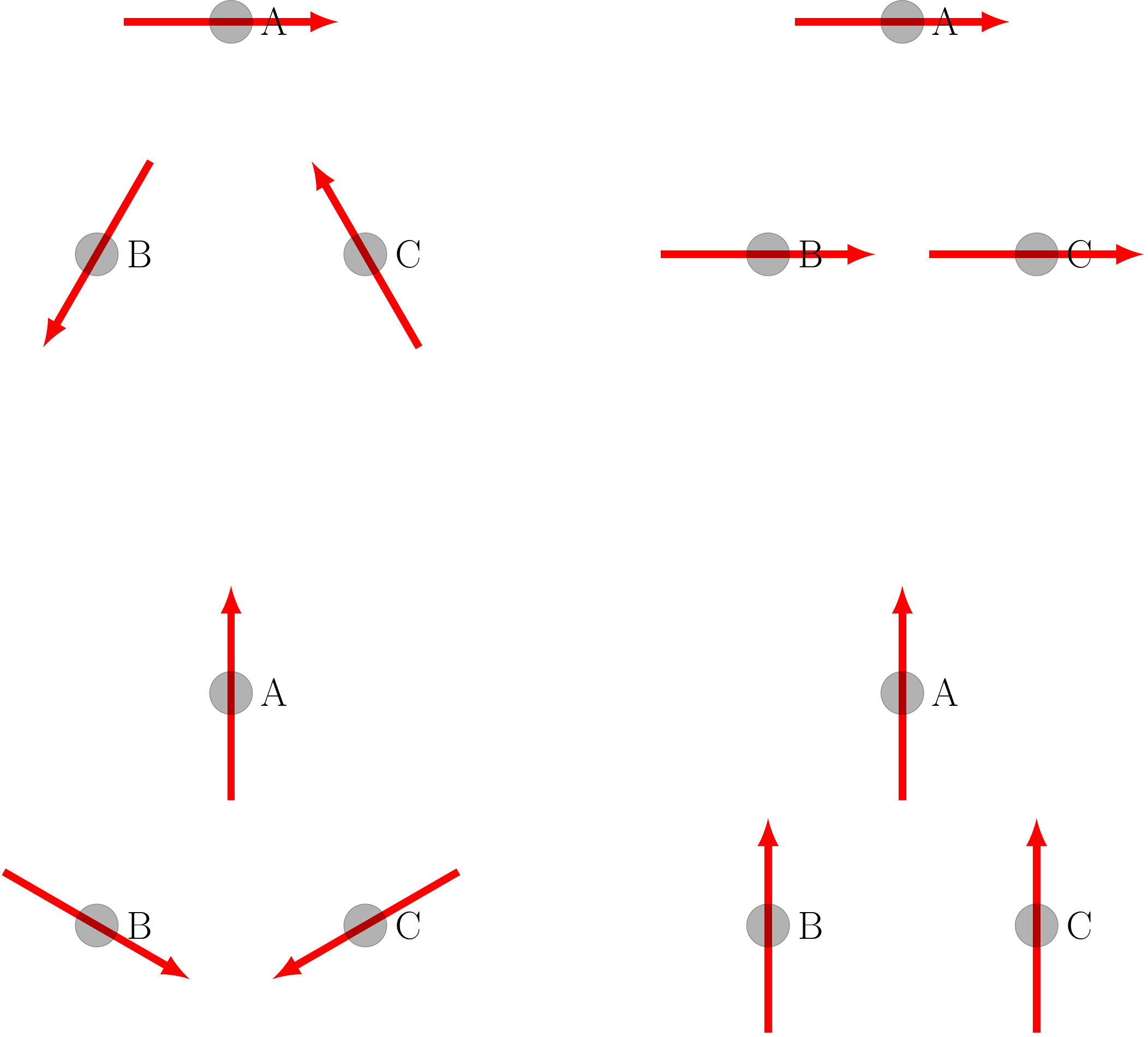}
	       \vspace*{0.2cm}
	    \end{minipage} \\ \colrule
    \end{tabular}
    \label{fig:irreps}
%\end{ruledtabular}
\caption{The irreducible representations corresponding to the low-energy spin configurations within the nine-dimensional subspace of the three Mn sublattices.
}
\end{figure}

%As discussed above the low-energy magnetic structure of the Mn$_3$Z compounds can be described with three sublattices each of them represented by a three-component spin vector. This means that the states of the model form a nine-dimensional space with the vectors set up from the sublattice spin-variables $\vec e = (\vec e_1, \vec e_2, \vec e_3)$.  
The representation $\mathbf R_g$ of the $D_{3h}$ point group on the nine-dimensional space defined within the three-sublattice model, see Eqs.\ \eqref{eq:Hcomp} and \eqref{eq:symmetry}, can be decomposed according to irreducible representations as 
\begin{equation}
\mathbf R_g = A^{\prime}_1 \oplus A^{\prime}_2 \oplus 2 E^{\prime} \oplus A^{\prime\prime}_2 \oplus E^{\prime\prime}\;,
\end{equation}
where $A^{\prime}_1$, $A^{\prime}_2$ and $A^{\prime\prime}_2$ are one-dimensional, while $E^{\prime}$ and $E^{\prime\prime}$ are two-dimensional irreducible representations. After projecting to irreducible subspaces we found that the low-energy chiral states and the ferromagnetic states correspond to the $A^{\prime}_1$, $A^{\prime}_2$ and the $2E^{\prime}$ irreducible representations as shown in Fig.~\ref{fig:irreps}. Here we distinguish the states $\Gamma_{3,x}$, $\Gamma_{3,y}$ and $\Gamma_{5,x}$, $\Gamma_{5,y}$ based on the orientation of the spin on the A sublattice. 

From Fig.~\ref{fig:irreps} we can conclude that the states $\Gamma_{3,x}$ and $\Gamma_{3,y}$ correspond to different one-dimensional irreducible representations, therefore, they are not degenerate by symmetry. So the energy of the $\Gamma_3$ states changes under in-plane global rotations. In contrast, the $\Gamma_{5,x}$ and $\Gamma_{5,y}$ states and also the FM$_x$ and FM$_y$ states form the basis of the same two-dimensional irreducible representation ($E^\prime$), thus, they are pairwise degenerate. This means that the $\Gamma_5$ and FM states are energetically insensitive to in-plane global rotations. Furthermore, sharing the same symmetry the $\Gamma_{5,x}$ and the FM$_x$ states are coupled and the same applies to the $\Gamma_{5,y}$ and the FM$_y$ states. In order to obtain the ground state of the model we have to diagonalize the Hamiltonian \eqref{eq:Hcomp} in the corresponding subspaces which leads to the diagonalization of the following matrix:
\begin{equation}
\left( 
\begin{array}{cc}
\vec{e}_{\Gamma_{5,\xi}} \;\mathbf J\;\vec{e}_{\Gamma_{5,\xi}}&
\vec{e}_{\Gamma_{5,\xi}} \;\mathbf J\;\vec{e}_{\text{FM}_\xi}\\
\vec{e}_{\text{FM}_\xi}\;\mathbf J\;\vec{e}_{\Gamma_{5,\xi}}&
\vec{e}_{\text{FM}_\xi}\;\mathbf J\;\vec{e}_{\text{FM}_\xi}
\end{array}
\right)
\end{equation}
where $\xi\in\{x,y\}$. The four resulting eigenstates can be cast into two degenerate weak ferromagnetic (WF) states that are slight distortions of the $\Gamma_5$ states and two denerate states that are modulations of the FM states. The low-energy weak ferromagnetic eigenstates can be written as
\begin{align}
\vec e_{\text{WF}_x} &=\mu  \, \vec e_{\Gamma_{5,x}}+ \nu \, \vec e_{\text{FM}_x} \label{eq:WFx} \\
\vec e_{\text{WF}_y} &=\mu  \, \vec e_{\Gamma_{5,y}}- \nu \, \vec e_{\text{FM}_y} \,,
\label{eq:WFy}
\end{align}
where $\mu^2+\nu^2=1$ and the explicit analytical expression for the $\nu$ parameter in terms of the sublattice model parameters indicates that
\begin{equation}
\nu \propto \left(J^x-J^y\right) + 2\left(K^x-K^y\right) \,.
\label{eq:nu}
\end{equation}

This result also suggests that the two different WF states possess a net magnetization of $\pm \nu$ coming from their FM component indicating that the  direction of this WF moment with respect to the A moment is different in the $x$ and $y$ cases.  Moreover, the weak ferromagnetic moment appears only if the $x$ and $y$ on-site and/or two-site anisotropy parameters differ from each other, but it does not occur if only the Dzyaloshinsky--Moriya interaction (DMI) is present in the system on top of the AFM isotropic interactions. Thus, in terms of group theoretical analysis, we regained the result of  Tomiyoshi and Yamaguchi \cite{tomiyoshi-1982} stating that the formation of the weak ferromagnetism in the Mn$_3$Sn compound happens due to magnetic anisotropy rather than DMI.  We note that the WF states \eqref{eq:WFx} and \eqref{eq:WFy} are stationary states of the Hamiltonian \eqref{eq:Hcomp} with the lowest energy, %what we can calculate directly but with group theoretical tools we reduced the diagonalization of the $9\times 9$ matrix to the diagonalization of 2 smaller ($2\times 2$) matrices and we get some physical insight into the nature of weak ferromagnetism. Unfortunately, this simple explanation of the weak ferromagnetic behavior goes into trouble, namely the resultant states 
but they do not refer to a set of spin vectors of unit length assumed in the classical Heisenberg model. In the next section we therefore revisit our investigation of the WF states within the space of classical spin states.

\subsection{Classical spin-model study}

As discussed in the previous section in terms of group-theoretical arguments, the energy of the $\Gamma_3$ state shows anisotropic behavior under global rotations around the $z$ axis, while the energy of the $\Gamma_5$ state is invariant to such rotations. This can be easily shown by calculating the rotational energies directly from     
Eq.\ \eqref{eq:H_sublatt} and using the parametric forms of the sublattice exchange matrices, Eqs.~\eqref{eq:JAA} and \eqref{eq:JBC}. For the $\Gamma_3$ state this gives
\begin{align}
E_{\Gamma_3}(\phi)=&\dfrac{3}{2}\left(J-\sqrt{3}D+\dfrac{3J^x-J^y}{2}-\dfrac{3K^x-K^y}{2}\right)\nonumber\\&- 3\left((J^x-J^y)-(K^x-K^y)\right)\sin^{2}\phi   \, ,
\label{eq:E_g3}
\end{align}
where  $\phi$  is the rotation angle around the $z$ axis with respect to the $y$ direction of the magnetic moment at the A atom (see the left panel of Fig.\ \ref{fig:spin_structure}). From the data of Table \ref{table:sublattice_params} it can be inferred that the in-plane anisotropy constant defined as the coefficient of the $\sin^2\phi$ term is negative, so the $\phi=0$ state ($\Gamma_{3,y}$) is the lowest in energy. For the energy of the $\Gamma_5$ states we get a constant indeed, \begin{equation}
E_{\Gamma_5} = \dfrac{3}{2}\left(J+\sqrt{3}D+\dfrac{J^x+J^y}{2}-\dfrac{K^x+K^y}{2}\right)\,.
\end{equation}
The energy difference between the $\Gamma_5$ and the $\Gamma_{3,y}$ states is then given by 
\begin{align}
E_{\Gamma_5}-E_{\Gamma_{3,y}} = 6\sqrt{3} D &-\dfrac{3}{2}(J^x-J^y-K^x+K^y) \, .
\label{eq:Ediff_G5-G3}
%\nonumber \\
%&+3(J^x-J^y-K^x+K^y) \sin^2(\phi) 
\end{align}
%where we can neglect the $\sin^2$ term if we consider the minimal energy $\Gamma_3$ state. 
Since for all considered Mn$_3$Z compounds the DMI is negative and is much larger in magnitude than the in-plane anisotropy term entering Eq.\ \eqref{eq:Ediff_G5-G3},  in each case the $\Gamma_5$ state has lower energy. This should be contrasted with the $L1_{2}$-type Mn$_3$Ir alloy, where the $\Gamma_3$ state is stabilized due to the magnetic anisotropy of about 10~meV \cite{szunyogh-2009}.

\begin{figure}[htb]                               
\includegraphics[width=0.9\linewidth]{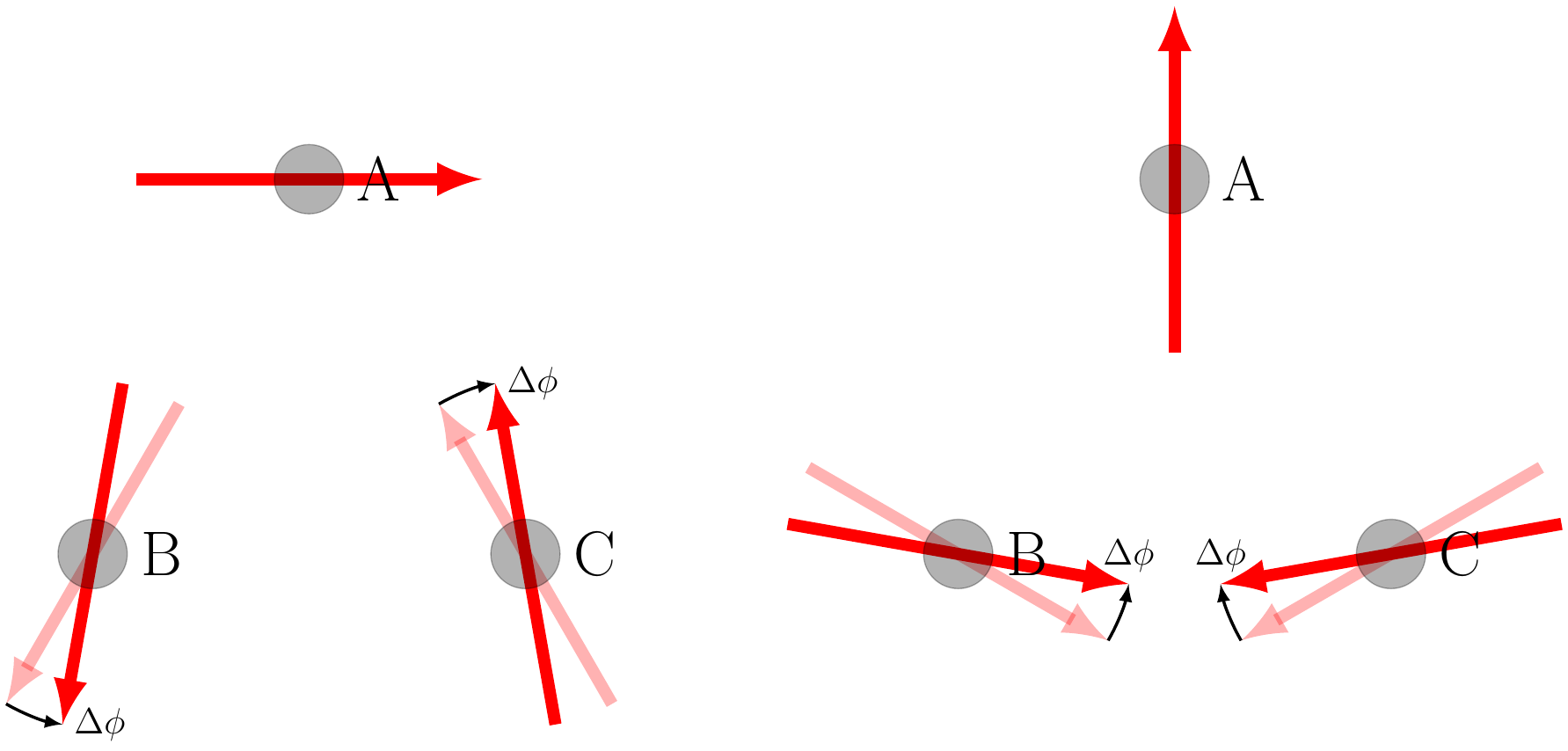}
\caption{Weak ferromagnetic distortions of the $\Gamma_5$ states: the WF$_x$ state on the left and the WF$_y$ state on the right. The shaded arrows show the spin directions in the original $\Gamma_5$ state. The distortion is parametrized by the tilting angle $\Delta\phi$.}
\label{fig:wf}
\end{figure}                                      

%The group-theoretical analysis indicated that the energy of the $\Gamma_5$ states can further be lowered upon weak ferromagnetic distortion.    
Within the classical spin model the weak ferromagnetic distortions can be parametrized by a tilting angle $\Delta\phi$ for two of the sublattices as illustrated in Fig.\ \ref{fig:wf}. %this way we stay on the space of the spin model states and 
The weak ferromagnetic moment is then related to $\Delta\phi$ as
\begin{equation}
m_{\text{WF}}(\Delta\phi) = 1 - 2\sin\left(\frac{\pi}{6}-\Delta\phi\right)\;.
\label{eq:wf_mom}
\end{equation}

\begin{figure}[htb]
\includegraphics[width=1\linewidth]{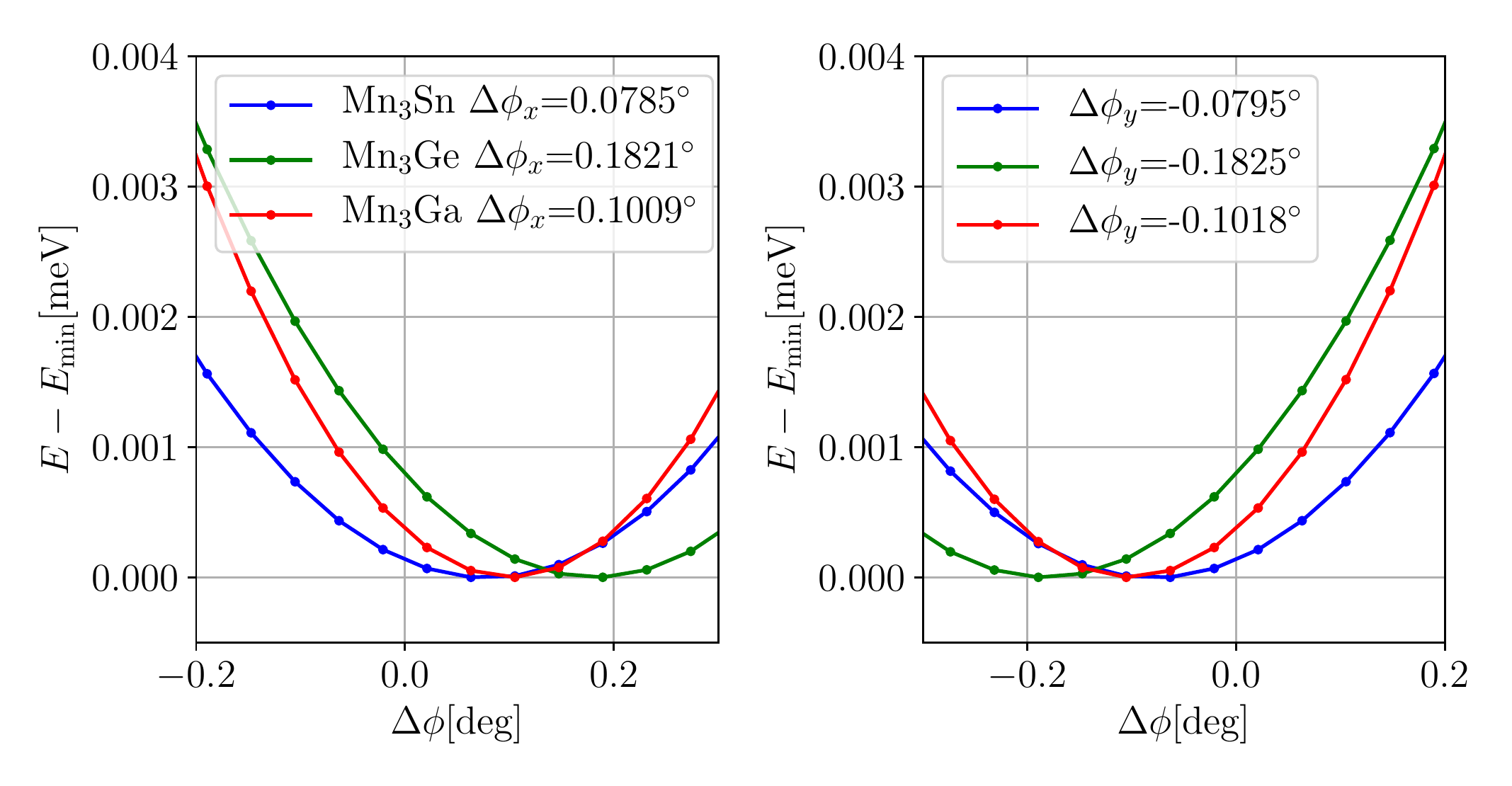}
\caption{Calculated energy curves for the Mn$_3$Z  compounds as a function of the $\Delta\phi$ weak ferromagnetic distortion angle. Left: WF$_x$ state, right: WF$_y$ state. The minimum positions $\Delta\phi_{x/y}$  were determined from parabolic fits.}
\label{fig:sm_wf}
\end{figure}

We calculated the energy of the weak ferromagnetic states as a function of the tilting angle $\Delta\phi$ based on the sublattice spin model. %Because of the small deviation from the $\Gamma_5$ state, in spirit of the magnetic force theorem \cite{mft} we attempted to estimate the energy differences in terms of bandenergies (single-particle energies), whereas the effective potentials and exchange fields were kept fixed as calculated self-consistently in the  $\Gamma_5$ state.  
The corresponding energy curves are shown in Fig.~\ref{fig:sm_wf} for both the WF$_x$ and  WF$_y$ distortions and for all the three compounds. As expected, for each alloy a clear parabolic minimum is obtained with
positive $\Delta\phi$ for WF$_x$ distortions and with negative $\Delta\phi$ for WF$_y$ distortions (cf.\ Fig.\ \ref{fig:wf}).
The weak ferromagnetic moments and the distortion angles obtained from the minima of the energy curves in Fig.~\ref{fig:sm_wf} are summarized in Table~\ref{table:sublattice_results}. Interestingly, the magnitudes of the WF$_x$ tilting angle are systematically smaller than those for WF$_y$, although the relative difference is about or less than 1~\%. Correspondingly, the size of the WF moments also somewhat differ for the two kinds of WF distortions, which contradicts the prediction of the group-theoretical analysis. It should be recalled again that the ground state obtained from group theory is outside the space of classical spin states. 

\begin{table}[htb]
    \caption{Calculated weak ferromagetic distortions and net magnetic moments for the Mn$_3$Z  compounds from the sublattice spin model with parameters obtained using the SCE-RDLM method.}
    \label{table:sublattice_results}
    \begin{ruledtabular}
        \begin{tabular}{c|cccc}
                   & $\Delta\phi_x$[deg] & $m_{\text{WF}_x}[10^{-3}]$ & $\Delta\phi_y$[deg] & $m_{\text{WF}_y}[10^{-3}]$ \\ \colrule
        Mn$_3$Sn   & 0.0785 & 2.37 & -0.0795 & -2.40 \\
        Mn$_3$Ge   & 0.1821 & 5.51 & -0.1825 & -5.51 \\
        Mn$_3$Ga   & 0.1009 & 3.05 & -0.1018 & -3.06 \\
        \end{tabular}
    \end{ruledtabular}
\end{table}

In order to gain deeper insight into the results obtained above and on the relationship of the WF distortion angle and the parameters of the spin model, we repeated the search for the minimum of the WF distortion energy analytically. The energy of the WF$_x$ distortion can be expressed as
\begin{align}
E_{\text{WF}_x}(\Delta\phi) &= 
  J \left( \cos(2\Delta\phi-\pi/3) + 2 \cos(\Delta\phi+\pi/3) \right) \nonumber \\  
&+ D \left( \sin(2\Delta\phi-\pi/3) + 2 \sin(\Delta\phi+\pi/3) \right) \nonumber \\
&+ \dfrac{J^x}{2} \left( \cos(2\Delta\phi-\pi/3) + 2\cos(\Delta\phi)- 1 \right)  \nonumber \\
&+ \dfrac{J^y}{2} \left( \cos(2\Delta\phi-\pi/3)-2\sqrt{3}\sin(\Delta\phi)+1 \right) \nonumber \\
&- (K^x-K^y) \cos(2\Delta\phi+\pi/3) - \dfrac{K^x}{4} - \dfrac{5K^y}{4}\;,
\label{eq:dphi_x}
\end{align}
\begin{comment}
\begin{align}
E(\Delta\phi_x) &= 
  J \left( \cos(2\Delta\phi_x-\pi/3) + 2 \cos(\Delta\phi_x+\pi/3) \right) \nonumber \\  
&+ D \left( \sin(2\Delta\phi_x-\pi/3) + 2 \sin(\Delta\phi_x+\pi/3) \right) \nonumber \\
&+ \dfrac{J^x}{2} \left( \cos(2\Delta\phi_x-\pi/3) + 2\cos(\Delta\phi_x)- 1 \right)  \nonumber \\
&+ \dfrac{J^y}{2} \left( \cos(2\Delta\phi_x-\pi/3)-2\sqrt{3}\sin(\Delta\phi_x)+1 \right) \nonumber \\
&- (K^x-K^y) \cos(2\Delta\phi_x+\pi/3) - \dfrac{K^x}{4} - \dfrac{5K^y}{4}\;,
\label{eq:dphi_x}
\end{align}
\end{comment}
and similar for $E_{\text{WF}_y}(\Delta\phi)$ by interchanging indices $x$ and $y$. 
%\begin{align}
%E(\Delta\phi_y) &= 
%  J \left( \cos(2\Delta\phi_y-\pi/3) + 2 \cos(\Delta\phi_y+\pi/3) \right)  \nonumber \\ 
%&+ D \left( \sin(2\Delta\phi_y-\pi/3) + 2 \sin(\Delta\phi_y+\pi/3) \right)  \nonumber \\
%&+ \dfrac{J^x}{2} \left( \cos(2\Delta\phi_y-\pi/3)-2\sqrt{3}\sin(\Delta\phi_y)+1 \right) \nonumber \\
%&+ \dfrac{J^y}{2} \left( \cos(2\Delta\phi_y-\pi/3)+2\cos(\Delta\phi_y) - 1 \right)   \nonumber \\
%&+ (K^x-K^y) \cos(2\Delta\phi_y+\pi/3) - 5K_x/4 - K_y/4\;.
%\label{eq:dphi_y}
%\end{align}
After expanding the above expression up to second order in $\Delta\phi$ it is easy to find its minimum yielding
\begin{equation}
\Delta\phi_{x} = -\dfrac{\sqrt{3}}{2}\dfrac{J^{x} - J^{y} + 2(K^{x} - K^{y})}{-3J - 3\sqrt{3}D - 2J^{x} - J^{y} + 2K^{x} - 2K^{y}}\;.
%\Delta\phi_{x/y} = -\dfrac{\sqrt{3}}{2}\dfrac{J^{x/y} - J^{y/x} + 2(K^{x/y} - K^{y/x})}{-3J - 3\sqrt{3}D - 2J^{x/y} - J^{y/x} + 2K^{x/y} - 2K^{y/x}}\;.
\label{eq:analytic_dphi}
%TODO solve overflow!
\end{equation}
and again similar for $\Delta\phi_y$  by interchanging the indices $x$ and $y$. 

This result has some important implications. Firstly, the numerator on the right-hand side of Eq.\ \eqref{eq:analytic_dphi} changes sign between the WF$_{x}$ and WF$_{y}$ distortions, while the denominator is always positive due to the large negative $J$ and $D$ parameters, which uniquely explains the sign change between the distortion angles $\Delta\phi_x$ and $\Delta\phi_y$. In addition, there is a small change in the denominator for the $x$ and $y$ cases, so we also found an analytic explanation for the deviation in the size of the corresponding tilting angles in Table~\ref{table:sublattice_results}. Secondly, the numerator is identical to the multiplicative factor we obtained from group theory for the weak ferromagnetic moment, see Eq.\ \eqref{eq:nu}. If we expand  Eq.~\eqref{eq:wf_mom} around $\Delta\phi=0$, we can see that for small distortions the weak ferromagnetic moment $m_\text{WF}$ is proportional to $\Delta\phi$ thus also with $J^{x} - J^{y} + 2(K^{x} - K^{y})$ as proposed by group theory.   %It means that in the formation of the weak ferromagnetic state the DMI don't play any role, it influences only the magnitude of the distortion via the denominator of Eq.\eqref{eq:analytic_dphi}. 
The weak ferromagnetism in the Mn$_3$Z  alloys is, therefore, qualitatively explained in the same way from group theory and from the classical spin model as being the consequence of nonzero on-site and/or two-site anisotropies,
$K^{x} - K^{y}$ and $J^{x} - J^{y}$, respectively. 

\subsection{Self-consistent calculations}

\begin{table*}[htb]
    \caption{The properties of the self-consistent weak ferromagnetic ground states of the Mn$_3$Z  compounds calculated from the SKKR method. The experimental results are taken from Refs.~\onlinecite{cable-1993}, \onlinecite{kadar-1971} and \onlinecite{kren-1970} for Mn$_3$Sn, Mn$_3$Ge and Mn$_3$Ga, respectively. The sign of the weak ferromagnetic moments refers to their orientation relative to the orientation of the moments in the A sublattice, see Fig.~\ref{fig:wf}.}
    \label{table:scf_results}
    \begin{ruledtabular}
        \begin{tabular}{c|ccc}
                      & Mn$_3$Sn & Mn$_3$Ge  & Mn$_3$Ga  \\ \colrule
        $\Delta\phi_x$& spin: -$0.006^\circ$ orbital: -$0.74^\circ$ & spin: $ 0.052^\circ$ orbital: $ 6.9^\circ$ & spin: $ 0.22^\circ$ orbital: $ 10.5^\circ$ \\
        $\Delta\phi_y$& spin: $ 0.010^\circ$ orbital: $ 0.78^\circ$ & spin: -$0.065^\circ$ orbital: -$5.9^\circ$ & spin: -$0.39^\circ$ orbital: -$8.9^\circ$\\
        m$_s^{\text{Mn}}[\mu_{B}]$ & 3.15 (3.17  $\pm$ 0.07 exp.) & 2.61 ($2.4\pm0.2$ exp.) & 2.60 ($2.4 \pm 0.2$ exp.) \\
        m$_{\text{WF}}[\mu_{B}]$ & x: -$0.003$, y: 0.003 (0.009 exp.)& x: 0.016, y: -$0.017$ (0.06 exp.)& x: 0.030, y: -$0.041$ (0.045 exp.)\\
        $E_{\text{WF}_x}-E_{\text{WF}_y}$[eV/f.u.] & $8.9\cdot 10^{-8}$ & -$3\cdot 10^{-9}$ & -$4.7\cdot 10^{-6}$ 
        \end{tabular}
    \end{ruledtabular}
\end{table*}

%The ground states of the Mn$_3$Z  compounds expected to be weak ferromagnetic $\Gamma_5$ states. 
We also performed self-consistent calculations to determine the ground states of the Mn$_3$Z  alloys by using the relativistic SKKR approach. We use the setup for the SKKR calculations discussed in Sec.~\ref{abinitio} and we let the magnetic moments relax from the $\Gamma_{5,x}$ and $\Gamma_{5,y}$ states to the corresponding weak ferromagnetic states. 
 This way we include the effect of the induced moment on the Z sites and also orbital-polarization effects into the calculations similarly as in the work of Sandratskii and K\"{u}bler\cite{sandratskii-1996} who investigated the weak ferromagnetism of Mn$_3$Sn from {\em ab initio} calculations. %Here we extend these studies to the other two compounds.

\begin{figure}[htb]
%\includegraphics[clip, trim=0cm 9.5cm 0cm 0cm, width=1\linewidth]{scf_wf.pdf}
%\vskip -10pt
%\includegraphics[clip, trim=0cm 0cm 0cm 12.4cm, width=1\linewidth]{scf_wf.pdf}
%\vskip -20pt
\includegraphics[width=\linewidth]{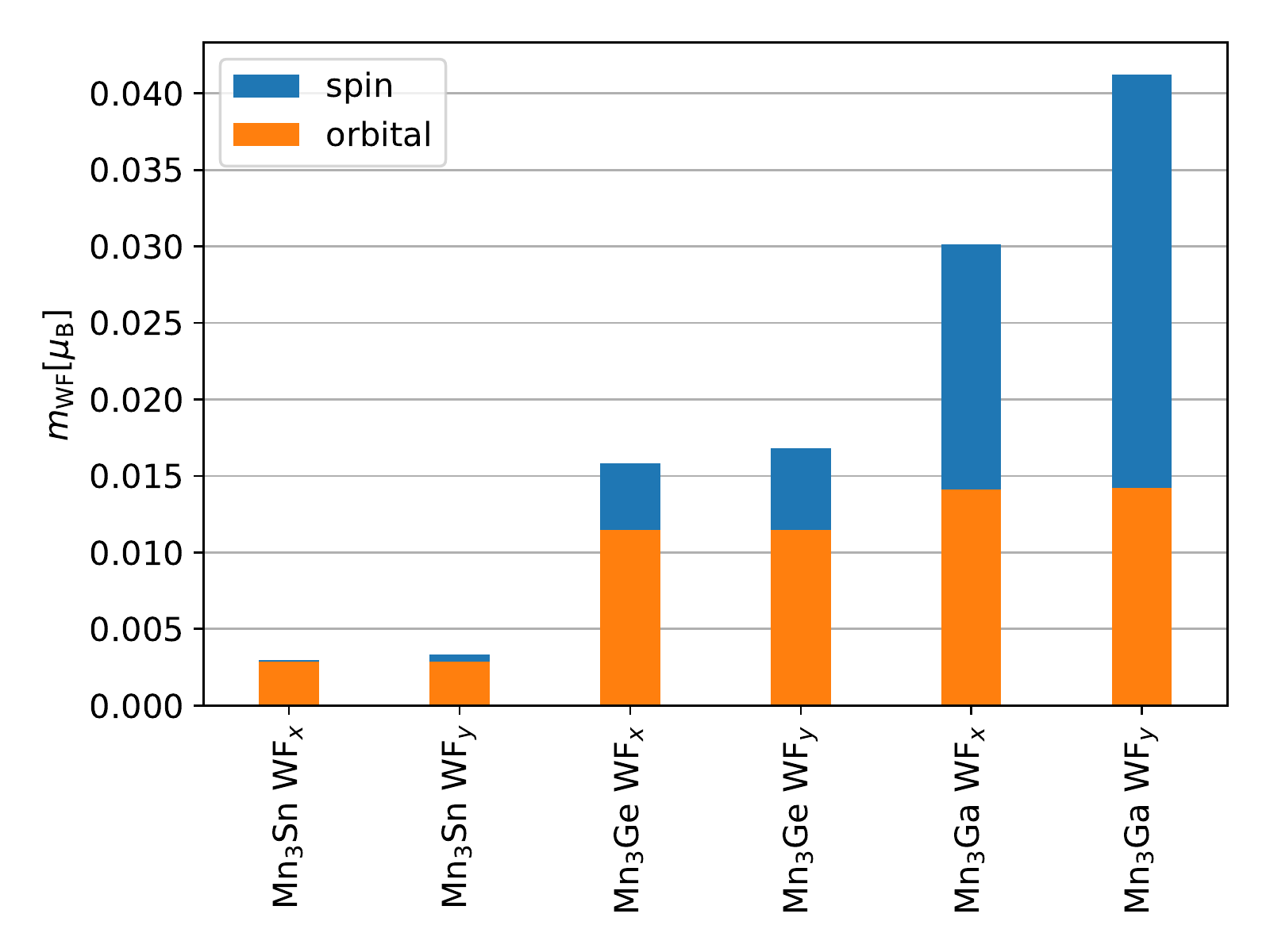}
\caption{Self-consistently calculated spin and orbital contributions to the weak ferromagnetic moments for the Mn$_3$Z  alloys. The moments are shown for both the WF$_x$ and WF$_y$ states.} 
\label{fig:scf_wf}
\end{figure}

In Table~\ref{table:scf_results} we summarize the main results  of the SKKR calculations for the weak ferromagnetic states of the Mn$_3$Z  compounds. 
First, we observe that the distortion of the spin vectors for Mn$_3$Sn is opposite for both the WF$_x$ and WF$_y$ states as proposed by the spin model, see Table~\ref{table:sublattice_results}. Regarding the expressions we derived for the distortion angles, Eq.~\eqref{eq:analytic_dphi}, this would mean that the self-consistent calculations predicted a magnetic anisotropy of opposite sign as compared to the spin model obtained from the SCE method. From Fig.~\ref{fig:scf_wf} presenting the spin and orbital contributions to the weak ferromagnetic moments we, however, see that in case of Mn$_3$Sn the WF moment is dominated by the orbital moment to which the spin model does not apply. In this case the tilting of spin moments seems to follow that of the orbital moments. Reassuringly, Sandratskii and K\"{u}bler\cite{sandratskii-1996} also obtained distortions of the same rotational sense for the WF$_x$ and WF$_y$ states,  resulting in a net moment antiparallel and parallel with the moment of the A atom, respectively. In contrast to our work, the spin-moment contribution reported in Ref.~\onlinecite{sandratskii-1996} is almost twice as large as the orbital contribution, though the total weak ferromagnetic moment, 0.004 $\mu_\text{B}$, is very close to our value (0.003 $\mu_\text{B}$). Remarkably, the distortion angle of the orbital moments is about two orders of magnitude larger than that of the spin moments, which, at least to a somewhat smaller extent, was also found in Ref.~\onlinecite{sandratskii-1996}.  It is worth to note that the SKKR codes rely on the solution of the Kohn--Sham--Dirac equation, while in Ref.~\onlinecite{sandratskii-1996}  the spin-orbit coupling is treated as an additive term to a scalar-relativistic Hamiltonian \cite{takeda-1978}.

For Mn$_3$Ge and Mn$_3$Ga, the self-consistent calculations yield weak ferromagnetic distortions of the same direction as found in the spin-model studies, although for Mn$_3$Ge the orbital contribution is still nearly three times larger than the orbital contribution, see Fig.~\ref{fig:scf_wf}. Only in case of Mn$_3$Ga, the spin-moment contributions become dominant and, curiously, this contribution shows a large difference between the WF$_x$ and WF$_y$ states which can not be understood based on the sublattice spin model. By contrast, the orbital contributions do not show this asymmetry for any of the systems under investigation, 
even though the quite enhanced distortion angles of the orbital moments for Mn$_3$Ge and Mn$_3$Ga display a remarkable anisotropy. 
As also indicated in Table~\ref{table:scf_results},  our {\em ab initio} calculations can not resolve with a reliable accuracy which of the two kinds of weak ferromagnetic states is energetically preferred. In case of Mn$_3$Ga, the WF$_x$ seems to be lower in energy, but from the very small energy difference of 4.7$\cdot10^{-6}$~eV/f.u.\ we rather conclude that the two weak ferromagnetic states are degenerate within the precision of the method we use.
  
A comparison with the experimental results is also shown in Table~\ref{table:scf_results}. The calculated spin moments of the Mn atoms are within the error range of the experiments. We note that the Mn moments slightly differ on the A and B (or C) sublattices as also reported in Ref.~\onlinecite{sandratskii-1996}.  The calculated weak ferromagnetic moments are also in the range of the experimental  values. The smallest value is obtained for Mn$_3$Sn as in the experiment, while for  Mn$_3$Ga we find a very good quantitative agreement with the measured value. The largest deviation from the experiment is found in the case of Mn$_3$Ge. A fair comparison between theory and experiment is, however, hardly possible for this alloy, since only off-stoichiometric samples could be prepared \cite{kadar-1971}, where Mn atoms can occupy Z positions leading to an enhanced net magnetic moment. Zhang \emph{et al.}\cite{zhang-2013}\ also performed density functional calculations using VASP\cite{vasp} for the hexagonal Mn$_3$Z alloys. As compared to their results, our calculated Mn spin moments are systematically larger by about 0.1 $\mu_\text{B}$, which might be attributed to the fact that in Ref. \onlinecite{zhang-2013} slightly smaller, optimized lattice constants and the generalized gradient approximation for the exchange-correlation functional \cite{PBE-1996} were used as opposed to the experimental lattice constants and the local density functional we employed, respectively. %They found the WF$_y$ state as ground state for all the three alloys and 
Reassuringly, however, they reported the weak ferromagnetic moments, 0.01 $\mu_\text{B}$ for Mn$_3$Sn and Mn$_3$Ge and 0.03 $\mu_\text{B}$ for Mn$_3$Ga, that are consistent with our values. 

As we noted already, the orbital moments have a significant weight in the weak ferromagnetic moment. Interestingly, the orbital moments of the Mn atoms show the behavior we found from group theory, namely that the weak ferromagnetic state can be decomposed as the linear combination of a $\Gamma_5$ and an FM state with the same mixing coefficients for the $x$ and $y$ state as given in Eqs.~\eqref{eq:WFx} and \eqref{eq:WFy}. In correspondence with these relationships, the net orbital moments for the WF$_x$ and WF$_y$ states are $m^\text{Mn}_{\ell} = 3\nu$ and $m^\text{Mn}_{\ell} = -3\nu$, respectively, while subtracting $\vec m^\text{Mn}_{\ell}/3$ from the orbital moments of each sublattice, a perfect $\Gamma_5$ state is obtained with local orbital moments of $\mu$.
The corresponding parameters for the three alloys are collected in Table~\ref{table:scp_orbital_states}. The opposite sign of $\nu$ for Mn$_3$Sn as compared to Mn$_3$Ge and Mn$_3$Ga reflects the opposite sign of the tilting angle as discussed above. Moreover, the increased magnitudes of $\nu/\mu$ for Mn$_3$Ge and Mn$_3$Ga correspond to the enhanced tilting angles for these alloys with respect to Mn$_3$Sn, see Table~\ref{table:scf_results}. Note that for Mn$_3$Ga
we found a slight deviation from the decomposition based on Eqs.~\eqref{eq:WFx} and \eqref{eq:WFy}. We believe that this impressive agreement between the distortion of the orbital moments and the group-theoretical prediction is due to the strongly non-rigid character of the orbital moments. The lack of a constraint of a constant magnitude allows the orbital moments to assume the superimposed WF configuration preferred by group theory.
\begin{table}[htb]
    \caption{The $\mu$ and $\nu$ parameters for the self-consistently calculated orbital moments, see Eqs.~\eqref{eq:WFx} and \eqref{eq:WFy},  using the convention where the $\Gamma_5$ and FM states are constructed from dimensionless unit vectors and the parameters are measured in units of $10^{-3}\mu_\text{B}$.}
    \label{table:scp_orbital_states}
    \begin{ruledtabular}
        \begin{tabular}{c|ccc}
               & Mn$_3$Sn & Mn$_3$Ge  & Mn$_3$Ga  \\ \colrule
          $\mu$&  40 & 33   & 26 \\
          $\nu$&  -0.6 & 4.3 & x: 5.07 \quad y: -5.10 
        \end{tabular}
    \end{ruledtabular}
\end{table}

%The great contribution of the orbital moments looks strange for the first sight because the orbital moment of the Mn atoms, what we can identify with the $\mu$ parameters in Table~\ref{table:scp_orbital_states}, are one magnitude smaller than the spin moments, but we found that they are much more distorted as shown on Fig.~\ref{fig:scf_wf} and in Table~\ref{table:scf_results}. 

%\subsubsection{The effect of the spin-orbit coupling}

Within the KKR formalism it is possible to scale down the spin-orbit coupling (SOC) selectively on different sites using a scalar relativistic framework for the treatment of the single-site scattering \cite{takeda-1978,ebert-1996}. In order to gain more insight into the effect of the spin-orbit coupling, we performed self-consistent calculations for each compound where we switched off the SOC either on the Z or on the Mn sites.  The results for the calculated spin and orbital parts of the net moments and their resolution into the Mn- and Z-atom contributions are summarized in Table~\ref{table:soc} together with the case where the SOC is included at all sites. 

\begin{table}[htb]
    \caption{Self-consistently calculated net magnetic moments of the weak ferromagnetic states in the Mn$_3$Z  alloys decomposed into spin and orbital contributions, as well as into contributions related to the Mn and Z atoms. All values are given in units of $10^{-3} \mu_\text{B}$. Both the WF$_x$ and WF$_y$ states are considered (see the second column). The cases when the SOC is included on all sites, only on the Mn sites and only on the Z sites, are indicated in the third column by `all', Mn and Z, respectively (Z=Sn, Ge, Ga).}
    \label{table:soc}
    \begin{ruledtabular}
        \begin{tabular}{ccc||rrr|rrr}
 &    & SOC & $m_s^{\text{Mn}}$ & $m_s^{\text{Z}}$ & $m_{s}$ & $m_\ell^{\text{Mn}}$ & $m_\ell^{\text{Z}}$ & $m_{\ell}$ \\ \hline
\multirow{6}{*}{\rotatebox[origin=c]{90}{Mn$_3$Sn}} & \multirow{3}{*}{\rotatebox[origin=c]{90}{WF$_x$}} & all & -0.13 & -0.01  & -0.14   & -1.80 & -1.04   &  -2.84 \\ 
                & & Mn  &  1.58 & -0.03  &  1.56   &  9.73 & -0.99  &   8.74 \\
                & & Sn  &  0.24 &  0.01  &  0.25   & -11.47 & -0.07 &  -11.53\\ \cline{2-9}
                & \multirow{3}{*}{\rotatebox[origin=c]{90}{WF$_y$}}& all &  0.49 &  0.01 &  0.49   &  1.80 &  1.04   &   2.85\\
                & & Mn  & -1.66 &  0.03  & -1.63   & -9.73 &  0.99  &  -8.74 \\
                & & Sn  & -0.24 & -0.01  & -0.25   &  11.47 &  0.07 &   11.53 \\\hline
\multirow{6}{*}{\rotatebox[origin=c]{90}{Mn$_3$Ge}} & \multirow{3}{*}{\rotatebox[origin=c]{90}{WF$_x$}}& all &  4.43 & -0.08  &  4.36   &  12.79 & -1.32   &  11.47 \\ 
                & & Mn  &  6.13 & -0.09  &  6.04   &  18.65 & -1.23   &  17.42 \\
                & & Ge  & -0.14 &  0.00 & -0.13   & -5.89 & -0.09 & -5.98 \\ \cline{2-9}
                & \multirow{3}{*}{\rotatebox[origin=c]{90}{WF$_y$}}& all & -5.39 &  0.10  & -5.30   & -12.81 &  1.32   & -11.49 \\
                & & Mn  & -7.98 &  0.11  & -7.86   & -18.69 &  1.23   & -17.45 \\
                & & Ge  &  0.11 & -0.00 &  0.11   &  5.89 &  0.09 &  5.98 \\\hline
\multirow{6}{*}{\rotatebox[origin=c]{90}{Mn$_3$Ga}} & \multirow{3}{*}{\rotatebox[origin=c]{90}{WF$_x$}}& all &  16.33 & -0.33   &  16.00   &  15.22 & -1.11   &  14.11 \\
                & & Mn  &  24.38 & -0.50   &  23.88   &  20.41 & -1.03   &  19.38 \\
                & & Ga  &  33.83 & -0.70   &  33.13   & -5.21 & -0.08 & -5.29 \\ \cline{2-9}
                & \multirow{3}{*}{\rotatebox[origin=c]{90}{WF$_y$}}& all & -27.57 &  0.55   & -27.02   & -15.33 &  1.11   & -14.22 \\
                & & Mn  & -38.97 &  0.77   & -38.20   & -20.57 &  1.03   & -19.54 \\
                & & Ga  & -21.69 &  0.43   & -21.26   &  5.21 &  0.08 &  5.29 \\
 
        \end{tabular}
    \end{ruledtabular}
\end{table}

%\begin{figure}[hbt]
%\includegraphics[width=1\linewidth]{soc.pdf}
%\caption{The effect of the SOC on the Mn and Z sites on the weak ferromagnetic ground state of Mn$_3$Z compounds. On the left we have SOC on each site, in the middle we turned off the SOC on the Mn sites and on the right we turn off the SOC on the Z sites.}
%\label{fig:soc}
%\end{figure}

First of all, Table~\ref{table:soc} indicates that net orbital moments induced by the SOC at the Mn and at the Z atoms add up almost perfectly when the SOC is switched on at each of the sites. Moreover, this is valid for the site-resolved contributions of the net orbital moments.  Apparently, such an additivity of the SOC induced net spin moments can not be established.  

In case of Mn$_3$Sn, the SOC at the Mn sites induces a sizable net spin moment mainly with contributions from the Mn atoms, but superimposed with the SOC of Sn this spin moment considerably reduces in size and it also changes sign. By contrast, the SOC of Sn raises an orbital moment of about 0.01 $\mu_\text{B}$ in size originating from the Mn atoms, which is, however, largely compensated by the orbital moment of opposite sign induced by the SOC of the Mn atoms. Although some orbital moment is induced also at the Sn sites being parallel to the orbital moment of the Mn atoms induced by the SOC of Sn, the total orbital moment remains parallel to that induced by the SOC of Mn. This peculiar cancellation of the orbital moments was also noticed and discussed by Sandratskii and K\"{u}bler\cite{sandratskii-1996}. 

In case of the Mn$_3$Ge and the Mn$_3$Ga alloys, it remains valid that both the net spin- and orbital moment mostly have contributions from the Mn atoms, while the Ge and Ga atoms add a negligible amount especially to the spin moment. 
The SOC of Mn plays a dominant role in the formation of the net moments in Mn$_3$Ge and this effect is only compensated in about 30~\% by the SOC of Ge. This explains the opposite direction of the weak ferromagnetic distortion, and correspondingly, the opposite direction of the net moment as compared to Mn$_3$Sn, where the effect of the SOC of Sn dominates.  The SOC of both the Mn and Ga atoms have a large effect in inducing a net spin moment in Mn$_3$Ga, but, interestingly, when switching on the SOC simultaneously on both sites, the total  spin moment is much less than the sum of the spin moments for selectively switched SOC. Moreover, the size of the spin moments induced by the SOC of Mn and Ga follow an opposite order for the WF$_x$ and WF$_y$ states, which might be connected with the large asymmetry of the spin moments for these weak ferromagnetic states in  Mn$_3$Ga.

%\subsection{The investigation of the induced moment}

\begin{figure}[htb]
\includegraphics[width=1\linewidth]{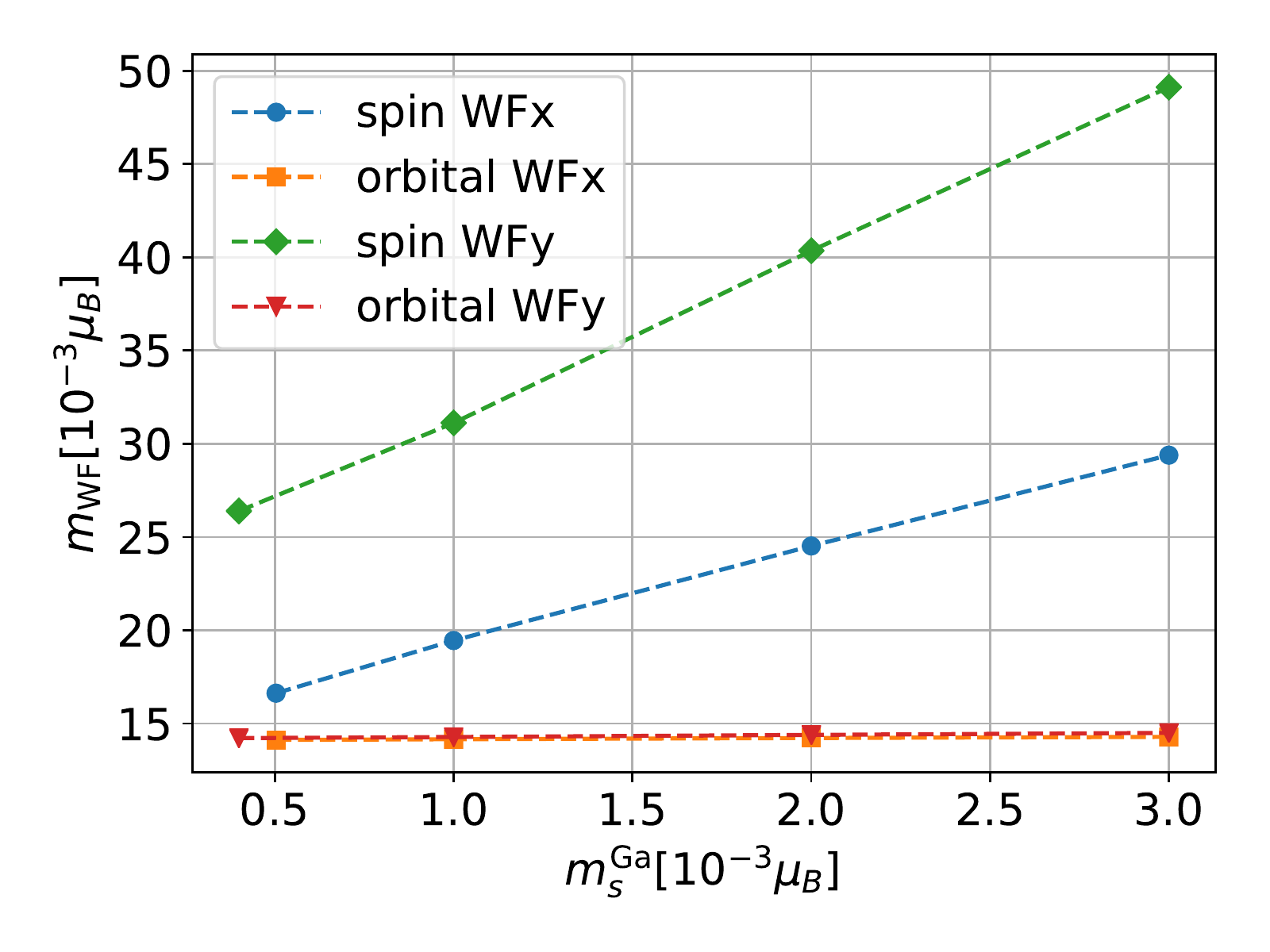}
\vskip -15pt
\caption{Magnitudes of the spin and the orbital contributions to the weak ferromagnetic moment in Mn$_3$Ga as a function of the induced spin moment at the Ga site in Mn$_3$Ga obtained from self-consistent constraining field  calculations.}
\label{fig:cf}
\end{figure}

The origin of the large asymmetry of the weak ferromagnetic spin moments for the WF$_x$ and WF$_y$ states obtained from the self-consistent density functional calculations is an unresolved issue. The largest anisotropy is found for Mn$_3$Ga, where also the largest induced spin moment at the Z atom is observed, likewise with a remarkably large asymmetry.  In order to see whether there is a connection between these two effects we performed self-consistent calculations by applying a longitudinal constraining field at the Ga site by which the spin moment of Ga could be set arbitrarily.  In Fig.~\ref{fig:cf} we plotted the net spin and orbital moments as a function of the induced spin moment on the Ga site. Or results indicate that the induced moment of Ga affects the net orbital moment only very moderately. By contrast, the net spin moments increase nearly linearly with the induced moment of Ga. The  corresponding lines for the WF$_x$ and WF$_y$ states have a different slope which implies that the asymmetry is also increasing with increasing induced moment of Ga. 
Extrapolating the lines to $m^\text{Ga}_s=0$ there still remains a difference, $m_{\text{WF}_y}- m_{\text{WF}_x} \simeq 10^{-2} \mu_\text{B}$, therefore, we can at best conclude that the induced moment of Ga is one of the sources of the asymmetry of the weak ferromagnetic moment in Mn$_3$Ga. The linear increase of the net spin moment with $m^\text{Ga}_s$ can qualitatively be understood in terms of an isotropic exchange coupling between the Mn and Ga spin moments, while the  asymmetry of the net spin moments can be attributed to the anisotropy, i.e to the tensorial nature of this coupling. 

%\bigskip

\section{Conclusions}

%We present combined spin model and first principles electronic structure calculations to study the weak ferromagnetism in bulk Mn$_3$Z  compounds. The spin model parameters were determined from a spin-cluster expansion technique based on the relativistic disordered local moment formalism implemented in the screened Korringa-Kohn-Rostoker method. We describe the magnetic ground state of the system within a three-sublattice model and investigate the formation of the weak ferromagnetic states in terms of the relevant model parameters. First, we give a group-theoretical argument how the point-group symmetry of the lattice leads to the formation of weak ferromagnetic states. Then we study the ground states of the classical spin model and derive analytical expressions for the weak ferromagnetic distortions by recovering the main results of the group-theoretical analysis. As a third approach we obtain the weak ferromagnetic ground states from self-consistent density functional calculations and compare our results with previous first principles calculations and with available experimental data. In order to trace the effect of the spin-orbit coupling at the Mn and Z sites on the weak ferromagnetic ground state we also perform self-consistent calculations by switching on the spin-orbit coupling on these sites selectively. In addtion, for the case of Mn$_3$Ga, we gain information on the role of the induced moment of Ga from constrained local density functional calculations.

We studied the weak ferromagnetism in the Mn$_3$Z (Z=Sn, Ge, Ga) alloys using a combination of \emph{ab initio} and spin model calculations. Using the point-group symmetry of the systems we set up a model for the three Mn sublattices including the relativistic terms of the Heisenberg Hamiltonian. The parameters of this model were obtained from \emph{ab initio} calculations relying on a spin-cluster expansion. Based on a group-theoretical analysis we showed that there are two degenerate ground states, WF$_x$ and WF$_y$, being the mixture of the $\Gamma_5$ and FM states. The analytical forms of the mixing coefficients imply that the weak ferromagnetic states will only form if the anisotropy parameters  distinguish between the $x$ and $y$ spin directions. We recovered this result from analytical expressions for the energy within the classical spin model. Our presented results are fully consistent with the original spin model description of weak ferromagnetism in Mn$_3$Sn by Tomiyoshi and Yamaguchi \cite{tomiyoshi-1982}.

We also performed self-consistent relativistic local density functional calculations in order to investigate the weak ferromagnetic states in a more involved way. In agreement with the seminal work for Mn$_3$Sn by Sandratskii and K\"{u}bler\cite{sandratskii-1996}, our results highlighted the significance of the orbital moments in the weak ferromagnetism of the Mn$_3$Z  alloys. A key observation from our calculations is that, as opposed to the spin moments, the orbital moments almost strictly follow the decomposition according to irreducible representations as predicted by group theory. As far as the spin moments of the Mn atoms and the weak ferromagnetic moments are concerned, we found a good agreement with the experiments for all the three alloys. The only exception is the net magnetic moment of Mn$_3$Ge, where presumably due to the off-stoichiometry of the samples the measured moment is significantly higher than the theoretical one.\cite{zhang-2013} By switching on the spin-orbit coupling at different sites selectively we discussed the role of the SOC on the formation of weak ferromagnetic moments and recovered the peculiar cancellation of orbital moments for Mn$_3$Sn reported in Ref.~\onlinecite{sandratskii-1996},  while we also found an argument for the opposite weak ferromagnetic distortion as compared to Mn$_3$Ge and Mn$_3$Ga.
By using constrained local density functional calculations 
 we established that the induced spin moment of Ga plays an important role in the formation of the weak ferromagnetic moment in Mn$_3$Ga and also it is one of the reasons for the observed large asymmetry of the net moment concerning the WF$_x$ and WF$_y$ states.

\begin{acknowledgments}
The authors are grateful for the financial support by the Hungarian National Scientific Research Fund (NKFIH) under project Nos.\ K115575 and PD124380, as well as by the BME Nanonotechnology FIKP grant of EMMI (BME FIKP-NAT). 
\end{acknowledgments}

%\appendix*
%\section{}\label{app:}

\end{document}